\documentclass[9pt,twocolumn,twoside]{osajnlArXiv}
\journal{ol} % Choose journal (ao, aop, josaa, josab, ol)
\graphicspath{{./figs/}}

\setboolean{shortarticle}{true} 
\title{Blind single-frame deconvolution by tangential iterative projections (TIP)}

\author[1,*]{Dean Wilding}
\author[1,2]{Oleg Soloviev}
\author[1]{Paolo Pozzi}
\author[1]{Carlas Smith}
\author[1,2]{Gleb Vdovin}
\author[1]{Michel Verhaegen}

\affil[1]{Delft Center for Systems and Control, Delft University of Technology, Mekelweg 2, 2628 CD Delft, The Netherlands}
\affil[2]{Flexible Optical B.V., Polakweg 10-11, 2288 GG Rijswijk, the Netherlands}

\affil[*]{Corresponding author: d.wilding@tudelft.nl}

\begin{abstract}
	Deconvolution serves as a computational means of removing the effect of optical aberrations from recorded images and is employed in many technical and scientific fields of study.  In most imaging scenarios the nature of the blurring kernel or point-spread function (PSF) of the imaging system is unknown and both the object and PSF can be estimated using different forms of mathematical optimisation.  The Tangential Iterative Projections (TIP) algorithm is a multi-frame deconvolution framework where multiple images can be combined to obtain a single estimate of the object.  It is shown here that this framework may be also used for single-frame deconvolution with a few modifications to the algorithm.  This step from multiple to one frame is non-trivial and greatly improves the applicability of the TIP framework to most imaging scenarios.
\end{abstract}

\setboolean{displaycopyright}{true}

\usepackage{todonotes}
\begin{document}
	
\maketitle

%% INTRODUCTION

A single-frame blind deconvolution is a ill-defined inverse problem that can be solved by the minimisation of the following general metric function, here one follows roughly after the approach by Sroubek \& Milanfar \cite{sroubek2012robust}:
\begin{equation}\label{eqn:metric}
\min_{h,o} \left|\left| i - h \ast o \right|\right|_2 + \lambda_h Q(h) + \lambda_o R(o),
\end{equation}
where $h,o,i \in \mathbb{R}^{N \times N}$ and represent the point-spread function (PSF), the object and the image respectively.  The functions $Q(h)$ and $R(o)$ are regularisation functions with their Lagrange multipliers $\lambda_h$ and $\lambda_o$ respectively and allow the solution of the optimisation to be steered to a realistic solution, as there are infinitely many combinations of $h$ and $o$ that may satisfy this minimisation for $\lambda_h=\lambda_o \equiv 0$.  For example, the trivial solution of the latter problem gives $h = \delta$ and $i = o$; however, since one knows \emph{a priori} that this is not the ground truth it is possible to constrain the optimisation to look for other compatible solutions.

Following Ayers and Dainty \cite{ayers1988iterative}, it is possible to do alternating ``projections'' \cite{NumericalRecipes2007} and solve this problem without explicit minimisation of the metric in Eq.~\ref{eqn:metric}.  Instead, a series of projections are made between allowed sets for the object and the PSF.  These projections constrain the solutions to the alternating minimisation of $o$ and $h$ to be real, non-negative and for $h$ to have specified spatial extent, also called limited support.  

It was shown in the authors' previous work Wilding \emph{et al.} \cite{wilding2017blind} that it was possible to use an alternating minimisation framework to elucidate both the point-spread functions (PSFs) and the object in a multi-frame blind deconvolution problem \cite{schulz1993multiframe}, this framework is called Tangential Iterative Projections (TIP).  The aforementioned choice of constraints has already known to provide a unique and noise-robust solution in this case \cite{Matson:09}.

The purpose of TIP was to be able to quickly and robustly deconvolve fluorescence microscopy images \cite{wilding2018pupil} with minimal noise amplification and using the minimal possible \emph{a priori} information to obtain reliable estimates of the ground truth, \emph{i.e.} the PSFs and fluorescence distribution that are really present.  Following this logic, the only tuning parameter in TIP was the support size of the PSFs.  The lack of \emph{a priori} information, however, meant that the TIP framework was not able to perform single-frame deconvolution.  Evidence of this failure was explicitly shown in the Discussion section of Ref.~\cite{wilding2017blind} and the purpose of this article is to propose modifications to the previously presented TIP framework to make it compatible with single-frame blind deconvolution problems.

\begin{figure}
	\centering
	\includegraphics[width=\columnwidth]{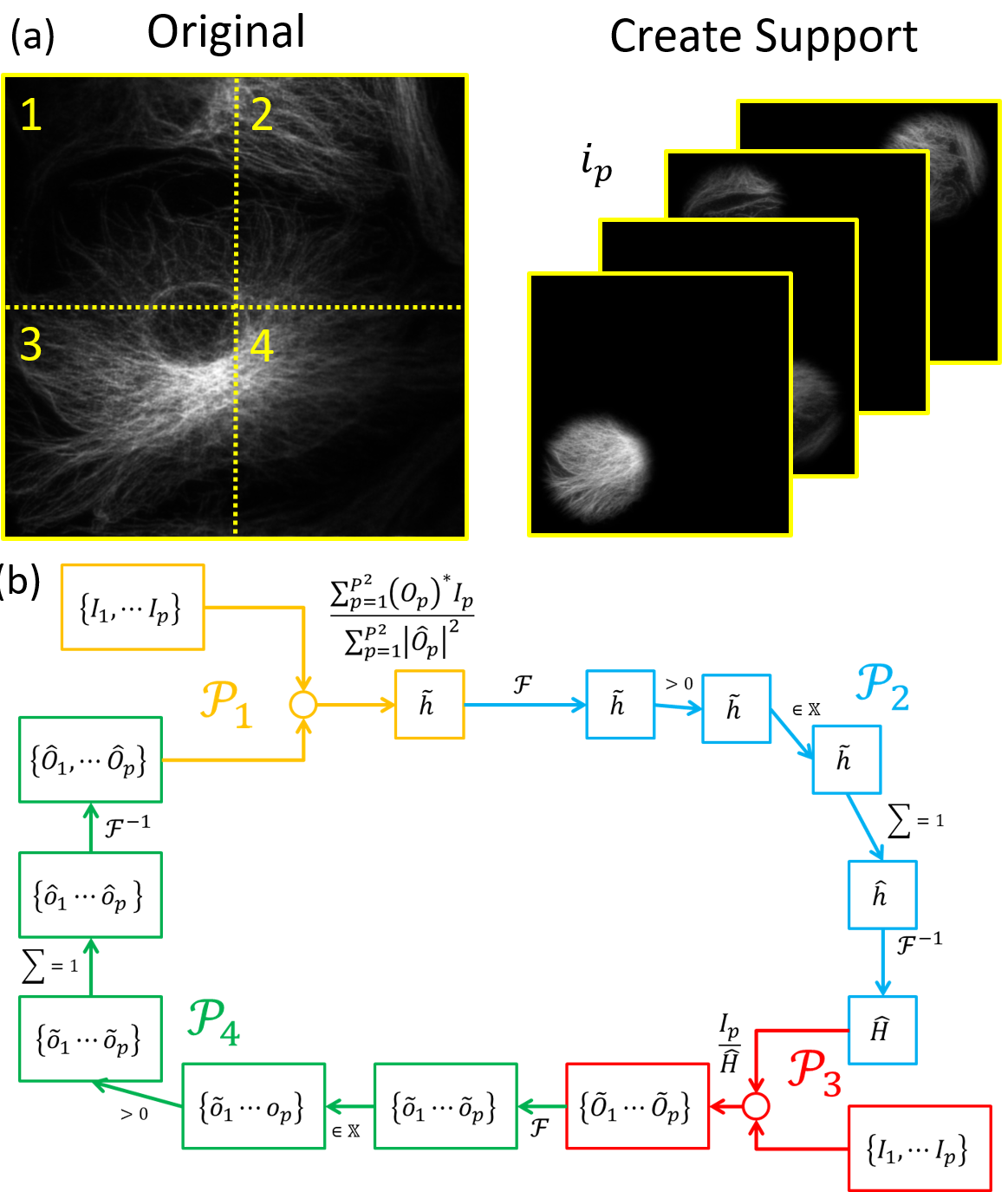}
	\caption{(a) An illustration of the process of converting the SF problem to a quasi-MF deconvolution problem. (b) The mathematical operations of the SF-TIP algorithm.}\label{fig:schema}
\end{figure}

The problem forthwith is addressed as a feasibility problem after Byrne \cite{byrne1998iterative} and the result of optimisation is that the solutions, $\hat{h}$ and $\hat{o}$, lie then inside their feasible sets.  Feasibility does not imply correctness and it is still possible that a solution whilst fitting inside the feasible set does not correspond to the ground truth.  These sets are named $\mathcal{H}$ for the PSF and $\mathcal{O}$ for the object and their properties are based on the physical parameters of incoherent imaging.  The mathematical operations involved in the process may be called projections: $\mathcal{P}_{1},\ldots,\mathcal{P}_{4}$ and are shown schematically in Fig.~\ref{fig:schema}(b).

%% MATHEMATICAL EXPLANATION

It may be seen that a multi-frame (MF) blind deconvolution problem is in fact actually a single-frame (SF) blind deconvolution problem with a special set of constraints on the object.  If one imagines that the set of images are taken in a temporal series, one acquires a three-dimensional dataset where the object is $\delta$-function in the temporal dimension or constant spectrally, \emph{i.e.} $(i_1,\ldots,i_N) = (o,0,\ldots,0) \ast (h_1,\ldots, h_N) = (o*h_1,\ldots,o* h_N)$.  This spectral constraint provides a finite-support constraint on the object and enables the MF problem to be solved with greater ease than the SF problem.

Likewise, it follows that if a MF problem can be converted to a SF problem, the same \emph{vice versa} is possible.  To solve the SF problem with the TIP algorithm, therefore, it is converted it to a MF problem.  This may be done in the following manner; firstly, the input image is split into a number of pseudo-patches $P \times P$.  These take the place of the multiple input images, they shall be labelled ${i}_p$ with corresponding object distribution ${o}_p$.  The isoplanatic case is considered and therefore, all these $P^2$ patches have the same PSF ${h}$.  The posed problem of Eq.~\ref{eqn:metric} in the Fourier domain with $x = \mathcal{F} \{ X \}$ becomes:
\begin{equation}\label{eqn:conversion}
\begin{aligned}
\min_{{H},{O}_p} & \hspace{0.2cm} \left|\left| \sum_{p=0}^{P^2}{I}_p - {H} \cdot {O}_p \right|\right|_2,\\
s.t. & \hspace{0.2cm} {h} \in \mathcal{H}  \\
	 & \hspace{0.2cm} {o}_p \in \mathcal{O}  \\
\end{aligned}
\end{equation}
In the original MF problem, the constraints applied to the object and PSF are the same.  There is a non-negativity constraint, a finite-support (for the object in the temporal dimension) and normalisation.  In the current case, one imagines that these patches are the ``temporal'' direction, as if one illuminated the object in sections.  In this way, the role of the object and PSF are reversed in the algorithm compared with Ref.~\cite{wilding2017blind}.  Since the PSF is now constant in ``time''.

One has now achieved a set of input images with the same PSF but different objects, this is however, not enough for good convergence properties.  There needs to be a finite-support constraint on the object, as with the MF problem.  The problem here is that for the majority of images there is no finite support on the object present.  It has been necessary, therefore, to find a method to introduce some finite support to the set $\mathcal{O}$.  A simple and effective way to do this is by introducing apodization and creating a non-binary mask ${m}_p$.  

In Fig.~\ref{fig:schema}(a) it may be seen how for an example image the SF can be decomposed into a set of images using these masks ${m}_p$.  The reconstruction process follows the process summarised diagrammatically in Fig.~\ref{fig:schema}(b), and may be seen to be similar to the outline given in Ref.~\cite{wilding2017blind}, with the variables $O$ and $H$ reversed and a new step introduced to explicitly apply the finite-support to the object.

The end goal of this process is to identity the PSF $h$ from the multiple observations, this can then be used to deconvolution the original single-frame to acquire the full object.  In SF problems noise-amplification at spatial frequencies where the OTF is small causes significant reconstruction artefacts.  Mathematically, when considering the spectra with additive noise $W$:
\begin{equation}
\begin{gathered}
I = H \cdot O + W, \\
O = \frac{I - W}{H}, \\
H \rightarrow 0,\ O = \frac{0 - W}{0} \rightarrow \infty,
\end{gathered}
\end{equation}
where the presence of zeros in the OTF are the biggest source of error in the reconstruction of the object.  When there are multiple OTFs this can be reduced by MF techniques, since the multi-frame linear deconvolution filter has the following form for $N$ independently acquired frames $I_n$:
\begin{equation}
O = \frac{\sum_{n=1}^{N} I_n \cdot \bar{H}_n }{\sum_{n=1}^{N} |H_n|^2 + \epsilon}, \\
\end{equation}
and in the presence of phase or amplitude diversity the $\sum_{n=1}^{N} |H_n|^2$ term in the denominator ensures that there is no division by small numbers, reducing the noise amplification.

When using multiple objects, however, it is not immediately obvious why this would help to condition this inverse operation.  One thing that is clear, is that it is not possible from the single-frame to reintroduce the information lost in the regions where $I$ and $H$ are close to zero.

Nevertheless, the main idea is that mathematically the problem of finding the object with multiple PSFs and finding the PSF based on multiple objects is in functionally equivalent, if the same constraints can be applied, therefore, by introducing an illumination constraint, here computationally introduced, the PSF can be observed multiple times.

The OTF is then given by:
\begin{equation}
H = \frac{\sum_{p=1}^{P^2} I_p \cdot \bar{O}_p }{\sum_{p=1}^{P^2} |O_p|^2 + \epsilon}, \\
\end{equation}
and the performance of the technique, therefore, depends on the ``diversity'' of the regions with the image, so that $\sum_{p=1}^{P^2} I_p \cdot \bar{O}_p \rightarrow 0 \Rightarrow \sum_{p=1}^{P^2} |O_p|^2 > 0$.  This arises due to the masking,  with $M_p = \mathcal{F}^{-1}\{m_p\}$, it can be seen that $I_p = M_p \ast I$ and for a good choice of mask $O_p \approx M_p \ast O$, this yields the following expression for the OTF:
\begin{equation}
H \approx \frac{\sum_{p=1}^{P^2} (M_p \ast I) \cdot (M_p \ast \bar{O}) }{\sum_{p=1}^{P^2} |M_p|^2 \ast |O|^2 + \epsilon}, \\
\end{equation}
and the convolution on the bottom of this expression allows the PSF to be recovered without noise amplification.  It is worth noting that acquiring frames with different illumination but a static aberration would be better suited for this procedure.

The object spectra $O_p$ are then found by the next projection:
\begin{equation}
O_p = \frac{I_p}{H} = \frac{M_p \ast I}{H}, \\
\end{equation}
and by iterating between these two projections, via the necessary constraints, a set of linear filters with mutual fidelity $1/\{O_p\}$ and $1/H$ may be elucidated, which have minimised the amount of energy outside of the support constraints.

%% TEST TARGET SIMULATION

The algorithm may be demonstrated first through a numerical simulation, so that it is possible to know with certainty what the object and the PSF are.  The case of isoplanatic imaging it is assumed the PSFs are generated by using a normally distributed set of 20 Zernike polynomials \cite{noll1976zernike} to give the simulated PSF. 

The USAF-1952 test target is used as an object and convolved with this PSF yielding an aberrated image, which is constrained to have a 16-bit image depth with no additive noise.  This image is then run through the SF-TIP framework for 100 iterations and the estimated object and PSF are output.

\begin{figure}
	\centering
	\includegraphics[width=\columnwidth]{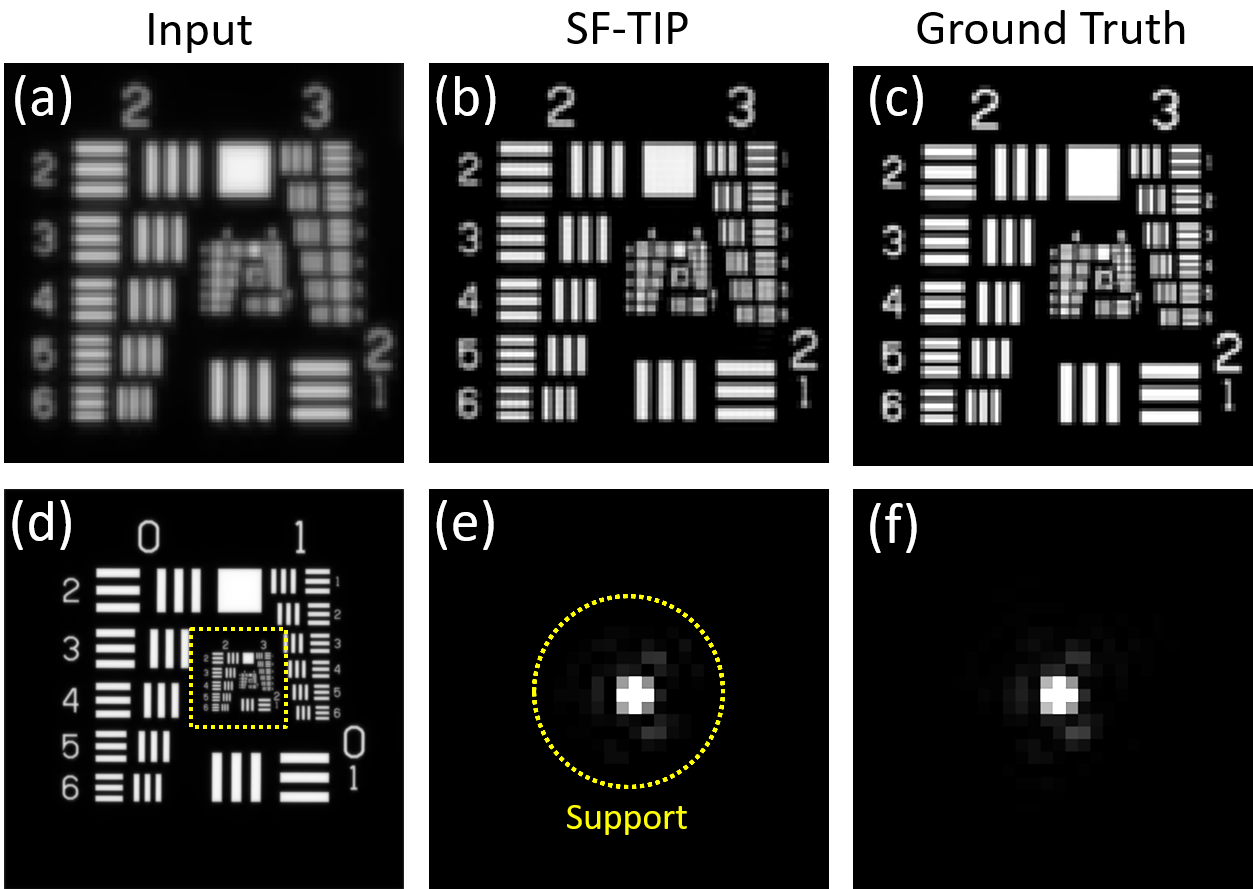}
	\caption{(a) Inset of (d) the aberrated input image. (b) Deconvolved object result using the PSF in (e) as retrieved by the SF-TIP algorithm. (c) Object used to generate the image.  (d) Full input image. (e) Reconstructed PSF. (f) Ground truth PSF.}\label{fig:testtarget}
\end{figure}

In Fig.~\ref{fig:testtarget} one sees the results of this test.  The top row (a)-(c) shows the central region-of-interest in the image given in full in (d), the ground truth object and PSF (c) and (f) are shown and can be compared with the retrieved (b) and (e).

Fig.~\ref{fig:testtarget}(a) shows that the image is aberrated and that resolution has been lost due to the convolution with the PSF in Fig.~\ref{fig:testtarget}(f).  The deconvolution result is given in Fig.~\ref{fig:testtarget}(b) and when it is compared with the ground truth object in Fig.~\ref{fig:testtarget}(c) one observes that the two are qualitatively similar.  The key points here are: the resolution of the original object has been restored, the aberration has been compensated for, and there are no strong numerical artefacts from the process.

If one now considers the PSF given by the deconvolution through SF-TIP, shown in Fig.~\ref{fig:testtarget}(e) and compares with the ground truth PSF given in Fig.~\ref{fig:testtarget}(f) one can conclude once again there is qualitative agreement.  The SF-TIP is able to identify the PSF used in a single-frame.

%% CONVERGENCE

To get a picture of the quantitative agreement one can look at the convergence properties of the algorithm.  This will be treated through numerical simulations rather than mathematical analysis.  A diverse set of aberrations, made from the set of Zernike modes, giving rise unique PSFs are convolved with an object source to give a set of images of 16-bit depth.

Each of these images are independently deconvolved using the process described in Fig~\ref{fig:schema}, and the Peak Signal-to-Noise Ratio (PSNR) is recorded at each step by comparing the current PSF estimate with the PSF used to generate the image, no additive noise is added to the images.  The PSNR is defined in the following manner for an image with $N$ pixels:

\begin{equation}\label{eqn:PSNR}
\mathrm{PSNR} = - 10 \log_{10} \left\{ \frac{1}{N} \sum |h - \hat{h}^{(k)}|^2   \right\}
\end{equation}

\begin{figure}
	\centering
	\includegraphics[width=\columnwidth]{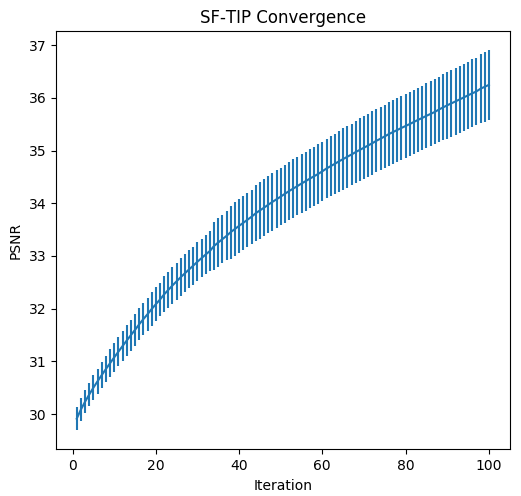}
	\caption{The mean PSNR of the PSF for 100 different image-PSF pairs deconvolved using the single-frame TIP framework.  The error bars show the standard deviation.}\label{fig:convergence}
\end{figure}

In Fig~\ref{fig:convergence} the results of this for 100 iterations of the algorithm are shown.  From this it can be concluded that for a particular image there is no certainty the algorithm will converge to the ground truth.  The general trends, however, can be observed by looking at these results statistically.  The algorithm will most of the time produce an improvement in the PSF PSNR between $5$--$7$ dB with the average being an $6$ dB increase.  Note here, this is not strictly speaking convergence in the sense of difference between successive iterations, \emph{i.e.} $|\hat{h}^{(k+1)} - \hat{h}^{(k)}|$, but instead a measure of the algorithms convergence to the ground truth.  The comparison of the PSF with the ground truth PSF is the best comparison for the algorithm performance, since information above the bandwidth is lost in the images.

%% EXPERIMENTAL RESULTS

The MF-TIP algorithm was shown to be successful in deconvolving image data from a variety of sources, here one of these sources shall be considered: horizontal imaging.  A 5 cm amateur telescope is used to acquire images through atmospheric turbulence on a day where the Fried parameter $r_0 < 5$ cm, \emph{i.e.} there are aberrations to correct in the images.

A region-of-interest (ROI) is displayed in (e) to (f) from the top images (a) to (c) respectively.  For comparison with an existing blind deconvolution method the algorithm from Kotera \emph{et al.} \cite{kotera2013blind} has been used because the MF-TIP algorithm was compared with a MF-variant and the code was available as a MATLAB code online; it shall be called MLE after Maximum Likelihood Estimation.  

In the results, one observes that both are able to perform the blind deconvolution, but the SF-TIP produces a sharper result with less filtering artefacts.  The results qualitatively look similar to those made in the MF-TIP article \cite{wilding2017blind} with MF-variant of Sroubek \& Milanfar \cite{sroubek2012robust}.  

\begin{figure}
	\centering
	\includegraphics[width=1.0\columnwidth]{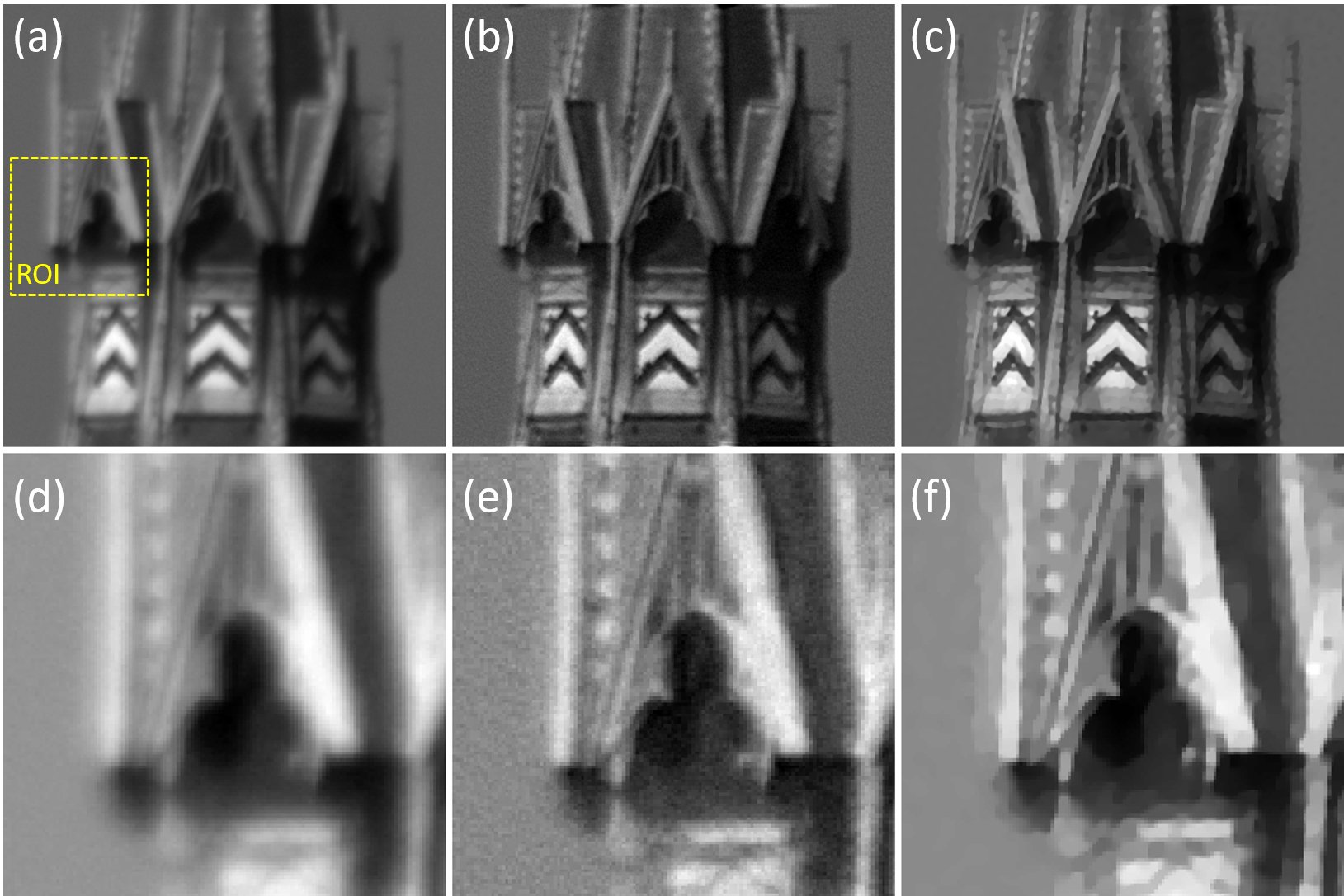}
	\caption{(a) Experimental frame from a horizontal telescope. (b) The SF-TIP deconvolved object. (c) The result using the algorithm from Kotera \emph{et al.} \cite{kotera2013blind}.   (d) A zoomed region of the experimental frame region-of-interest (ROI) shown in (a). (e) The same region as (b) from the SF-TIP deconvolved object. (f) The same region as (c) from the Kotera \emph{et al.} \cite{kotera2013blind} deconvolved object.}\label{fig:tower}
\end{figure}

Whilst in the experimental case it is impossible to compare the ground truth, it is possible to note that the image has gotten sharper and there are details present that where not present in the original image.  There also does not appear to be any significant artefacts from the deconvolution process, e.g. ringing.

%% DISCUSSION

From the numerical simulations, the convergence analysis and experimental testing, it is possible to conclude that SF-TIP is able to solve the blind deconvolution problem states in Eq.~\ref{eqn:metric}.  The step from using multiple frames to a single frame has not been trivial step the elucidation of the schema in Fig.~\ref{fig:schema}, deemed to be the pivotal breakthrough.

Whilst the TIP framework has been shown to work for the single-frame case, it should be noted there are fundamental limitations to this type of deconvolution.  Namely, it is impossible with one measurement to make an empirical observation of the ground truth.  In the multi-frame case, the TIP framework is able to extract the information from the diverse observations --- where the best result is obtained when the sum OTF spectrum has no zeros --- \emph{i.e.} all information has been transmitted.  On the other hand, in the single-frame case this missing information is inferred via the assumed \emph{a priori}.

The SF-TIP procedure is not found to require more iterations to extract the PSF and the object from the image than is seen with MF-TIP.  Since multiple frames are used the computational load of SF-TIP is similar to the MF-variant.

Moreover, it may be concluded that under certain situations the advantages of single-frame deconvolution negate the downsides of less reliability.  Fundamentally, there is an implicit or explicit cost associated with the acquisition of frames in many disciplines and reducing this to the absolute minimum is therefore, a major positive step.  Secondly, multiple frames are only of any benefit if it is possible to perturb the OTF of the system, or if there is naturally occurring diversity such as turbulence.  Single-frame deconvolution does not require this added diversity element and therefore, it is applicable in all imaging scenarios and situations.

With that said, the nature of TIP in general is that it does not perform well on sparse images, such as star fields, this is because there is not enough information in the spectrum to separate $H$ from $O$.  Given in SF-TIP that one uses the assumption that the image spectrum has a particular composition, this enhances the problem and therefore, SF-TIP should only be used on extended source images, such as microscopy or horizontal imaging.

To conclude this letter, in the author's previous work a MF blind deconvolution algorithm called TIP was presented.  It was shown in this work that it could not deconvolve SF blind deconvolution problems due to lack of \emph{a priori} information.  In this letter, a methodology that allows the TIP framework to be successfully applied to single-frame images has been demonstrated with numerical and experimental results.  This greatly widens the applicability of the framework developed in Wilding \emph{et al.} \cite{wilding2017blind} allowing it be applied in any imaging system for the compensation of aberrations when imaging extended sources.

\section*{Funding}
ERC Grant Agreement No. 339681.

\section*{Acknowledgements}
The authors would like to acknowledge the technical contributions of W.J.M. van Geest and C.J. Slinkman.

% Bibliography
\bibliography{refs}

\begin{thebibliography}{10}
\newcommand{\enquote}[1]{``#1''}

\bibitem{sroubek2012robust}
F.~Sroubek and P.~Milanfar, {\protect\JournalTitle{IEEE Transactions on Image
  Processing}} \textbf{21}, 1687 (2012).

\bibitem{ayers1988iterative}
G.~Ayers and J.~C. Dainty, {\protect\JournalTitle{Optics letters}} \textbf{13},
  547 (1988).

\bibitem{NumericalRecipes2007}
W.~H. Press, S.~A. Teukolsky, W.~T. Vetterling, and B.~P. Flannery,
  \emph{Numerical Recipes 3rd Edition: The Art of Scientific Computing}
  (Cambridge University Press, 2007).

\bibitem{wilding2017blind}
D.~Wilding, O.~Soloviev, P.~Pozzi, G.~Vdovin, and M.~Verhaegen,
  {\protect\JournalTitle{Optics Express}} \textbf{25}, 32305 (2017).

\bibitem{schulz1993multiframe}
T.~J. Schulz, {\protect\JournalTitle{JOSA A}} \textbf{10}, 1064 (1993).

\bibitem{Matson:09}
C.~L. Matson, K.~Borelli, S.~Jefferies, J.~Charles C.~Beckner, E.~K. Hege, and
  M.~Lloyd-Hart, {\protect\JournalTitle{Appl. Opt.}} \textbf{48}, A75 (2009).

\bibitem{wilding2018pupil}
D.~Wilding, P.~Pozzi, O.~Soloviev, G.~Vdovin, and M.~Verhaegen,
  {\protect\JournalTitle{Opt. Express}} \textbf{26}, 14832 (2018).

\bibitem{byrne1998iterative}
C.~Byrne, {\protect\JournalTitle{Inverse Problems}} \textbf{14}, 1455 (1998).

\bibitem{noll1976zernike}
R.~J. Noll, {\protect\JournalTitle{JOsA}} \textbf{66}, 207 (1976).

\bibitem{kotera2013blind}
J.~Kotera, F.~{\v{S}}roubek, and P.~Milanfar, \enquote{Blind deconvolution
  using alternating maximum a posteriori estimation with heavy-tailed priors,}
  in \emph{International Conference on Computer Analysis of Images and
  Patterns,}  (Springer, 2013), pp. 59--66.

\end{thebibliography}

\bibliographyfullrefs{refs}

\end{document}